# DeepEMC-T2 Mapping: Deep Learning-Enabled T2 Mapping Based on Echo Modulation Curve Modeling

Haoyang Pei[1,2,3], Timothy M. Shepherd[1,2], Yao Wang[3], Fang Liu[4,5], Daniel K Sodickson[1,2], Noam Ben-Eliezer[1,2,6,7], Li Feng[1,2]

1 Bernard and Irene Schwartz Center for Biomedical Imaging, Department of Radiology, New York University Grossman School of Medicine, New York, NY, USA.

2 Center for Advanced Imaging Innovation and Research (CAI$^2$R), Department of Radiology, New York University Grossman School of Medicine, USA

3 Department of Electrical and Computer Engineering and Department of Biomedical Engineering, NYU Tandon School of Engineering, New York, NY, USA

4 Athinoula A. Martinos Center for Biomedical Imaging, Massachusetts General Hospital, Charlestown, Massachusetts, USA

5 Harvard Medical School, Boston, Massachusetts, USA

6 Department of Biomedical Engineering, Tel Aviv University, Tel Aviv, Israel

7 Sagol School of Neuroscience, Tel Aviv University, Tel-Aviv, Israel

**Word Count:** 4661

**Running Header: DeepEMC-T2 Mapping**

**Grant Support:** NIH (P41EB017183, R01EB030549, R21EB032917, R21EB031185, R01AR081344, and R01AR079442)

Address correspondence to:

Li Feng, PhD

Center for Advanced Imaging Innovation and Research (CAI$^2$R)

New York University Grossman School of Medicine

660 First Avenue

New York, NY, USA 10029

Email: lifeng.mri@gmail.com

## Abstract

**Purpose**: Echo modulation curve (EMC) modeling can provide accurate and reproducible quantification of T2 relaxation times. The standard EMC-T2 mapping framework, however, requires sufficient echoes and cumbersome pixel-wise dictionary-matching steps. This work proposes a deep learning version of EMC-T2 mapping, called DeepEMC-T2 mapping, to efficiently estimate accurate T2 maps from fewer echoes without a dictionary.

**Methods**: DeepEMC-T2 mapping was developed using a modified U-Net to estimate both T2 and Proton Density (PD) maps directly from multi-echo spin-echo (MESE) images. The modified U-Net employs several new features to improve the accuracy of T2/PD estimation. MESE datasets from 68 subjects were used for training and evaluation of the DeepEMC-T2 mapping technique. Multiple experiments were conducted to evaluate the impact of the proposed new features on DeepEMC-T2 mapping.

**Results:** DeepEMC-T2 mapping achieved T2 estimation errors ranging from 3%-12% in different T2 ranges and 0.8%-1.7% for PD estimation with 10/7/5/3 echoes, which yielded more accurate parameter estimation than standard EMC-T2 mapping. The new features proposed in DeepEMC-T2 mapping enabled improved parameter estimation. The use of a larger echo spacing with fewer echoes can maintain the accuracy of T2 and PD estimations while reducing the number of 180-degree refocusing pulses.

**Conclusions:** DeepEMC-T2 mapping enables simplified, efficient, and accurate T2 quantification directly from MESE images without a time-consuming dictionary-matching step and requires fewer echoes. This allows for increased volumetric coverage and/or decreased SAR by reducing the number of 180-degree refocusing pulses.

## Introduction

Quantitative T2 mapping has great potential to provide additional information of clinical value to the current diagnostic MRI workflow for various applications (1), including the detection of brain lesions (2), assessment of myocardial edema (3, 4), diagnosis of biochemical changes in the knee cartilage (5, 6), detection, characterization, and grading of assorted lesions in the body (7-10), and muscle physiology research (11). Multi-echo spin-echo (MESE) imaging is a common MRI acquisition strategy for T2 mapping. However, accurate quantification of T2 relaxation in MESE imaging presents significant challenges in clinical practice, including long scan times, limited volumetric coverage, high specific absorption rate (SAR), and the inherent bias of T2 estimation caused by B1+ inhomogeneity and imperfect slice profiles that create stimulated and indirect echoes in the MESE echo train (12).

To reduce estimation bias, T2 maps can be estimated from MESE images using Bloch simulation-based signal matching (12). With this technique, stepwise Bloch simulations of the experimental pulse sequence are first performed to create a database or dictionary of underlying signal evolution, each corresponding to a specific combination of T2 and B1 values. Due to B1+ inhomogeneity and imperfect slice profiles, each simulated T2-decaying signal in the dictionary follows a generalized echo-modulation curve (EMC) instead of a pure exponential decay. A T2 map is generated by matching the MESE images with the pre-generated dictionary. Specifically, the signal evolution of each pixel in the MESE images is compared to every entry in the dictionary to identify the best match. The T2 value corresponding to the best-matched dictionary entry is then allocated to the respective pixel location. Repeating this process for each pixel location generates a T2 parameter map. Based on the estimated T2 map, a proton density (PD) map can then be computed by back-projecting the first echo image using an exponential model (12). This EMC-T2 mapping technique enables accurate T2 mapping in clinically-feasible scan times (13-18). It can be performed as a post-processing step using DICOM images directly reconstructed from the MESE acquisition on the scanner. Therefore, it does not require modification of the imaging sequence. To accelerate the dictionary matching process, EMC-T2 mapping was also optimized to incorporate a gradient-descent search algorithm, as described in reference (19).

While EMC-T2 mapping exhibits improved accuracy compared to exponential model-based T2 fitting, it faces several significant challenges. First, pixel-wise parameter matching in EMC-T2 mapping requires postprocessing time and computational resources that may limit widespread clinical translation (12). Despite the significant acceleration of the mapping procedure achieved by the latest algorithm (19), it still demands a substantial computational cost. Second, robust T2 estimation requires an adequate number of echoes for dictionary matching, which necessitates a long echo train length (ETL) in MESE imaging. This results in increased SAR and also limits the number of slices that can be acquired within each TR. Reducing the number of echoes, as will be seen in our results presented below, can lead to significant T2 estimation error.

In this study, we propose a deep learning version of EMC-T2 mapping, referred to as DeepEMC-T2 mapping, to address these challenges. The training of our DeepEMC-T2 mapping model is based on a modified U-Net structure incorporating several novel components, including a data consistency loss, spatiotemporal filters, and the removal of the Pooling/Downsampling and Upsampling layers. After network training, DeepEMC-T2 mapping enables efficient and accurate estimation of both T2 and PD maps directly from a reduced number of echoes compared to standard EMC-T2 mapping and does not require a dictionary). We also assessed the performance and quality of DeepEMC-T2 with different numbers of source echoes. In the following sections, we first review the standard EMC-T2 mapping technique. The main framework of our proposed DeepEMC-T2 mapping technique is then described, followed by an evaluation of its performance in multiple in-vivo imaging experiments.

## Methods

### Recap of Standard EMC-T2 Mapping

In an ideal situation with a perfect RF slice and B1+ profile, the signal evolution along the echo train in MESE imaging follows an exponential decay based on the following equation:

$$M(t) = PD \cdot e^{-\frac{t}{T_2}} \qquad [1]$$

Here, $M$ represents the signal intensity at different echo times, $PD$ denotes the proton density and $t$ denotes the echo times. In practice, B1+ inhomogeneity, particularly at high field strengths, can lead to stimulated and indirect echoes, which cause substantial signal contamination in acquired images. This contamination occurs as a result of the separation of magnetization into three coherence pathways with each refocusing pulse (12). In this case, the resulting T2 decay of MESE no longer follows a pure exponential model as shown in Equation 1, although this model is commonly employed for T2 fitting in clinical practice.

The overall framework of EMC-T2 mapping to address these limitations is illustrated in Figure1a. Based on the MESE imaging parameters used for data acquisition, stepwise Bloch simulation is performed to simulate all coherence pathways, including stimulated and indirect echoes, throughout the echo train. The simulation is repeated for a pre-defined range of T2 and B1+ values to create a dictionary of theoretical EMCs. Each entry in the dictionary represents a single echo-modulation curve that is associated with a unique combination of a T2 value and a B1+ value. For a fixed imaging protocol, the simulation needs to be performed only once as a preprocessing step.

The estimation of a T2 map in EMC T2 mapping involves matching the signal evolution of each pixel in the acquired MESE images to the simulated EMC dictionary. Specifically, for each pixel location, the L2 norm of the difference between the signal evolution and each entry of the EMC dictionary is calculated. The T2 value associated with the dictionary entry that has the minimal L2 norm with the acquired signal is then assigned to the current pixel location. This matching process is repeated for all pixel locations in all acquired image slices to generate multi-slice T2 maps. Based on the estimated T2 map, a PD map can then be computed for each slice by back-projecting the first echo MESE image to time t=0 using the following equation:

$$PD_r = I_{r(t=TE1)} \; / \; e^{-\frac{TE1}{T2_r}} \qquad [2]$$

Here, $I_{(t=TE1)}$ represents the first echo image, TE1 is the first echo time and T2 is the estimated T2 map. Note that an exponential model can be used to generate a PD map since the MESE signal undergoes pure exponential decay between the excitation and first acquisition event (20), and is thus not contaminated by stimulated and indirect echoes.

## DeepEMC-T2 Mapping: The Network Design

DeepEMC-T2 mapping employs deep neural networks to achieve simplified, more efficient, and accurate T2 estimation directly from MESE images. The training of the DeepEMC-T2 network follows a supervised scheme, where reference T2 and PD maps are estimated from MESE images with 10 echoes using the dictionary-based matching implemented in the standard EMC-T2 method.

The DeepEMC-T2 network is implemented using the U-Net with several important modifications. First, the standard U-Net (21) employs spatial filters that may not be optimal for extracting the temporal features of T2 decay information. Instead, spatiotemporal filters are implemented to explore image features along both the spatial and temporal dimensions in MESE images. The spatial branch of spatiotemporal filters is important in exploring the relationship between T2 values and tissue types and can contribute to improved estimation accuracy. The modified U-Net structure (shown in Supporting Information Figure S1) consists of an encoder and a decoder with skip connections between them. The linear transforms are employed at the final layer to map the features at each pixel to T2 and PD maps following all spatiotemporal encoder and decoder blocks, which is implemented using two 2D convolutional layers (Conv2D) with a kernel size of $1 \times 1$. To estimate a consistent value range of the output T2 maps, the output activation function for predicting T2 maps was modified as a negative Log-Sigmoid to extend the output value range to $[0, \infty]$.

Second, the standard U-Net usually incorporates Pooling and Upsampling layers to enlarge the spatial receptive field while concurrently reducing computational costs. The spatial receptive field is defined as the area of the input image that a specific neuron in a convolutional layer considers or takes into consideration while making predictions or extracting features. A larger spatial receptive field proves beneficial for tasks requiring predictions based on large regions of input images, such as segmentation and object detection. For the EMC-T2 mapping, however, a larger spatial receptive field could potentially introduce irrelevant features to the prediction of T2 values at a specific pixel location because the estimation of T2 values at each pixel location in the conventional EMC-T2 relies solely on the temporal decay of a given spatial location in the MESE images. Accordingly, we propose to eliminate the standard Pooling and Upsampling layers in the modified U-Net used in DeepEMC-T2. This aims to reduce the spatial receptive field, ensuring that the prediction of the T2 value at a specific location relies on the temporal decay of the given spatial location and its nearby spatial locations in the MESE images.

Third, a data-consistency (DC) loss is incorporated into network training. Specifically, the training of the DeepEMC-T2 network enforces two types of loss functions, including (1) an L1 loss between the predicted T2 and PD maps (denoted as $\mathrm{T2_p}$ and $\mathrm{PD_p}$) and the reference T2 and PD maps (denoted as $\mathrm{T2_r}$ and $\mathrm{PD_r}$) and (2) two additional loss functions to minimize the L1 norm of the difference between the first-echo reference image (denoted as $\mathrm{I_{1r}}$ from the reference MESE images) and a synthesized first-echo image that is computed through forward-projection according to the predicted T2 and PD maps and the reference T2 and PD maps, as shown in Figure 1b. The loss function is formulated as Equation 3 below.

$$\mathrm{L}_\theta = ||\mathrm{T2_r} - \mathrm{T2_p}||_1 + ||\mathrm{PD_r} - \mathrm{PD_p}||_1 +$$

$$\lambda_1 ||\mathrm{PD_r} e^{-\frac{\mathrm{TE_1}}{\mathrm{T2_p}}} - \mathrm{I_{1r}}||_1 + \lambda_2 ||\mathrm{PD_p} e^{-\frac{\mathrm{TE_1}}{\mathrm{T2_r}}} - \mathrm{I_{1r}}||_1 \qquad [3]$$

Note that for data consistency, two synthesized first-echo images are generated by using $\mathrm{T2_p/PD_r}$ and $\mathrm{T2_r/PD_p}$, respectively, corresponding to the two loss functions for data consistency as shown in Equation 3 and Figure 1b. $\lambda_1$ and $\lambda_2$ are empirically set to 1 to balance the weights of each part of loss in this work.

The performance of our modified U-Net combining the three features above to improve DeepEMC-T2 mapping was demonstrated in different experiments as described below.

**Datasets**

A total of 68 MESE datasets were retrospectively collected for our study. The datasets were acquired from 38 healthy subjects (8 males, 6 females [mean age=43.5±11.2 years] and 24 datasets with anonymized subject information) and 30 patients with multiple sclerosis (MS) (8 males and 22 females, mean age=48.8±9.0 years) on two 3T clinical MRI scanners (MAGNETOM Skyra and TimTrio, Siemens Healthineers, Erlangen, Germany) using a vendor-provided MESE sequence. All subjects provided written consent forms prior to the MRI scans. Each dataset includes 26 slices with 10 echoes each. Other imaging parameters included: FOV=220x206mm$^2$, matrix size=128x120, slice thickness=3mm, TR=4100ms, first TE=15ms, $\Delta$TE=15ms. Data acquisition was accelerated using parallel imaging (GRAPPA: GeneRalized Auto-calibrating Partial Parallel Acquisition) with an acceleration rate of 2. All images were directly reconstructed on the scanners. 46 datasets acquired in 33 healthy controls and 13 MS patients were used for training, 7 datasets acquired in 5 healthy controls and 2 MS patients were used for validation, and evaluation was performed in the rest 15 datasets that were all acquired in MS patients.

**Training Configuration**

The model weights in network training were updated using the adaptive gradient descent optimization (ADAM) algorithm (22) with a learning rate of 0.0003. A batch training strategy was implemented with a minibatch size of 4-16 depending on the number of echoes selected from MESE images for training. The total iteration steps were 200 epochs, and the best model was chosen when the lowest L1 loss between reference and estimated T2 maps was achieved in the validation datasets. The training was performed using PyTorch (version 2.0) on a server with an NVIDIA Tesla A100 GPU card. All experiments follow the above training configuration.

**Evaluation**

A total of six experiments were designed to evaluate our DeepEMC-T2 mapping technique. Experiment 1 compared DeepEMC-T2 mapping with standard EMC-T2 mapping (simply referred to as EMC-T2 mapping hereafter) for estimating T2 and PD maps. Experiment 2 evaluated whether the DC loss could improve the accuracy of parameter estimation in DeepEMC-T2 mapping. Experiment 3 assessed whether removing the Pooling and Upsampling layers in U-Net could enhance the accuracy of parameter estimation in DeepEMC-T2 mapping. Experiment 4 investigated whether replacing the spatial filters with spatiotemporal filters in U-Net could improve the accuracy of parameter estimation in DeepEMC-T2 mapping. The spatial filter blocks were built by removing the temporal branch of the spatiotemporal filter blocks.

In the first four experiments, the comparison used MESE images generated with all 10 echoes, the first 7 echoes (echo 1-7), the first 5 echoes (echo 1-5), and the first 3 echoes (echo 1-3). MESE images with fewer echoes were generated by truncating late echoes. For DeepEMC-T2 mapping, the truncated images were used both for training and inference, and the T2 and PD maps from all the 10 echoes estimated using standard EMC-T2 were used as ground truth references for supervised training.

In addition, Experiment 5 evaluated the impact of echo spacing in MESE images on the accuracy of parameter estimation for both EMC-T2 and DeepEMC-T2 mapping. Specifically, this was performed using MESE images with the first 3 echoes (echo 1-3), the first 5 echoes (echo 1-5), and selected 3 echoes (echo 1,3,5). Note that the echo spacing is increased when using echo 1,3,5 in MESE imaging for parameter estimation.

Experiment 6 aimed to assess the performance of DeepEMC-T2 mapping with a fully connected network (FCN, also known as a multilayer perceptron) instead of the U-Net. This experiment was designed as FCN requires less or no training datasets and has been widely utilized in quantitative MRI applications, including MR fingerprinting (23). In this scenario, the FCN was only trained on the decay curve at each pixel location of MESE images to estimate the T2 and PD values at corresponding pixel location, as shown in Figure S2. It did not take the entire MESE images as input, and therefore, the estimation of T2 and PD maps did not depend on any spatial information or image features. The hidden layer of FCN employed ReLU as the activation function. In addition, the activation function of the output node for predicting T2 values was modified to a negative Log-Sigmoid. Similar to the experiments above, FCN-based DeepEMC-T2 mapping was trained using decay curves of the MESE images from all 10 echoes, 7 echoes (echo 1-7), 5 echoes (echo 1-5), and 3 echoes (echo 1-3). The loss function was the same as that used in the U-Net-based DeepEMC-T2 mapping. The model weights in the FCN training were updated using the ADAM algorithm with a learning rate of 0.0001. A batch training strategy was implemented with a minibatch size of 1024.

For all the six experiments, the differences between the predicted T2 maps and the reference T2 maps were assessed by calculating the pixel-wise relative error in different T2 ranges from 40-160ms with an increment of 40ms. For example, 40-80ms was set as T2 range 1, 80-120ms was set as T2 range 2, etc. Different masks representing different T2 ranges were generated from the ground truth T2 maps, and the obtained masks were applied to the predicted T2 maps directly without adaption for calculating the pixel-wise relative error. The pixel-wise relative error was calculated as:

$$\mathrm{Error}(\%) = \frac{|\mathrm{Pred} - \mathrm{Ref}|}{\mathrm{Ref}} \times 100\% \qquad [4]$$

The differences between the predicted PD maps and the reference PD maps were also assessed by calculating the pixel-wise relative error averaged over all pixel locations. Paired student t-test was used for statistical analysis, where a P value less than 0.05 was considered statistical significance.

## Results

### Experiment 1: Comparison Between DeepEMC-T2 and Standard EMC-T2 Mapping

Figure 2a and Figure 3a show a representative case comparing T2 and PD maps estimated using EMC-T2 mapping and DeepEMC-T2 mapping from MESE images with different numbers of echoes. The T2 and PD maps estimated from standard EMC-T2 mapping using all 10 echoes were treated as the reference for calculating the error maps. DeepEMC-T2 mapping consistently yielded improved accuracy compared to EMC-T2 mapping across different numbers of reduced echoes.

Figure 2b summarizes the quantitative comparison of T2 estimation averaged over all the testing cases (n=15). The improvement of DeepEMC-T2 mapping over EMC-T2 mapping reached statistical significance (P<0.05) for all the T2 ranges across different numbers of reduced echoes. Similarly, Figure 3b provides a quantitative overview of PD estimation averaged over all the testing datasets. The improvement of DeepEMC-T2 mapping over EMC-T2 mapping reached statistical significance (P<0.05) for 5 and 3 echoes. These results highlight that DeepEMC-T2 mapping enables more accurate parameter estimation than EMC-T2 mapping when the number of echoes is reduced.

### Experiments 2-4: The Impact of DC Loss, Removing Pooling and Upsampling layers, and Spatiotemporal Filters

Figure 4 shows a representative case comparing T2 and PD maps estimated using DeepEMC-T2 mapping with and without a DC loss. The error maps indicate that incorporating a DC loss into DeepEMC-T2 training can slightly improve the estimation of both T2 and PD maps, as indicated by the red arrows. The quantitative comparison in all the testing cases is summarized in the first and second rows of Table 1. The improvement from incorporating a DC loss reached statistical significance ($P<0.05$) only in one range of T2 estimation (40-80ms) and in PD estimation for 7 and 3 echoes as highlighted by the red stars in the second row of Table 1.

Figure 5 presents a representative case comparing T2 and PD maps estimated using DeepEMC-T2 mapping with and without the Pooling and Upsampling layers. The error maps highlight a notable reduction in error for both T2 and PD estimation by removing the Pooling and Upsampling layers. The quantitative comparison in all the testing cases is summarized in the first and third rows of Table 1. The improvement of removing Pooling and Upsampling layers reached statistical significance ($P<0.05$) for all ranges of T2 estimation and PD estimation across different numbers of echoes.

Figure 6 compares T2 and PD maps estimated using DeepEMC-T2 mapping with spatial and spatiotemporal convolution kernels for one case. Error maps visually demonstrate that including spatiotemporal convolution filters in DeepEMC-T2 mapping improves the estimation of both T2 and PD. The quantitative analysis of pixel-wise relative error in different T2 ranges for all the testing cases, as shown in the first and fourth rows of Table 1. The improvement reached statistical significance ($P<0.05$) for all ranges of T2 estimation and PD estimation across different numbers of echoes.

**Experiment 5: The Impact of Echo Spacing**

Figure 7a and Figure 8a show a representative example comparing T2 and PD maps estimated from MESE images with echo 1,3,5 with a larger echo spacing and with 5 echoes (echo 1-5) and 3 echoes (echo 1-3) with the default echo spacing using both EMC-T2 and DeepEMC-T2 mapping. It can be seen that increasing the echo spacing while reducing the number of echoes has a neglectable impact on both T2 and PD estimation for both EMC-T2 and DeepEMC-T2 mapping.

Figure 7b and Figure 8b summarize the corresponding quantitative comparison for T2 and PD estimation in all the testing cases. For both EMC-T2 and DeepEMC-T2 mapping, there was no significant difference between echo 1,3,5 with a larger echo spacing and echo 1-5 with the default echo spacing for all ranges of T2 estimation (P>0.05). The errors for PD estimation from echo 1,3,5 and echo 1-5 were all below 2% for both EMC-T2 and DeepEMC-T2 mapping. This confirms our visual observation that using a larger echo spacing with a reduced number of echoes can maintain the accuracy of parameter estimation. Besides, there is a significant improvement for both T2 and PD estimation (P<0.05) from 3 echoes (echo 1-3) with the default echo spacing to echo 1,3,5 with a larger echo spacing. This suggests that increasing the echo spacing while maintaining the same number of echoes can improve the accuracy of parameter estimation. The improvement of DeepEMC-T2 mapping over EMC-T2 mapping also reached statistical significance (P<0.05) for all ranges of T2 estimation and PD estimation across different numbers of echoes.

**Experiment 6: Comparison Between U-Net and FCN in DeepEMC-T2 mapping**

Figure 9 shows a representative case comparing T2 and PD maps estimated using DeepEMC-T2 mapping with FCN and our modified U-Net with different numbers of echoes. It is obvious that the modified U-Net enabled improved accuracy compared to FCN across different numbers of echoes. The quantitative comparison in all the testing cases is summarized in the first and last rows of Table 1. The improvement reached statistical significance (P<0.05) for all ranges of T2 estimation across different numbers of echoes and for PD estimation with 7 and 3 echoes.

## Discussion

Deep learning has shown great promise for improving quantitative MRI in general (24). It can be used to improve quantitative MRI reconstruction (24-27), to improve the estimation of the parameters from reconstructed images (28-31), and to accelerate more complex multicomponent analysis (18). In this work, we developed DeepEMC-T2 mapping, a deep learning framework for T2 quantification based on EMC modeling using the MESE sequence. After network training, DeepEMC-T2 mapping enables direct and efficient estimation of both T2 and PD maps without a time-consuming dictionary matching step that is required in standard EMC-T2 mapping. Moreover, DeepEMC-T2 mapping allows for a more accurate estimation of both T2 and PD from a reduced number of echoes compared to EMC-T2 mapping. The DeepEMC-T2 network incorporated several novel components to improve performance, and the impact and effectiveness of these new features were investigated in different experiments. This study also demonstrated that the echo spacing in MESE imaging can affect the accuracy of DeepEMC-T2 and EMC-T2 mapping.

**Accuracy and Performance of DeepEMC-T2 Mapping**

DeepEMC-T2 mapping enables accurate estimation of both T2 and PD from all 10 echoes compared to the reference standard. The error was below 3% for T2 estimation in different T2 ranges relevant to the in vivo brain and was below 0.8% for PD estimation. Meanwhile, when the number of echoes was reduced (e.g., from 10 echoes to 7, 5, and 3 echoes by truncating the late echoes), DeepEMC-T2 Mapping yielded more accurate parameter estimation than EMC-T2 mapping. This is because DeepEMC-T2 mapping can more effectively estimate the underlying T2 decay patterns with a reduced number of echoes, while conventional dictionary matching becomes less robust and reliable with fewer echoes. The TR of a MESE sequence is typically on the order of seconds, including the time required to acquire an echo train and some additional idle time to ensure signal recovery at the end of each echo train. To improve imaging efficiency, the idle time for a given image slice can usually be used to acquire other slices in an interleaved manner. As a result, reducing the number of echoes can help reduce the SAR of MESE acquisition and potentially increase the total number of image slices.

**New Features of DeepEMC-T2 Mapping**

Three new components have been incorporated into the DeepEMC-T2 mapping framework to improve the accuracy of parameter estimation, including a DC loss, removal of the Pooling and Upsampling layers that are typically implemented in the standard U-Net, and spatiotemporal convolution kernels. First, the incorporation of the DC loss increases the accuracy of parameter estimations, albeit with a minor improvement (Figure 4 and Table 1). Second, eliminating the Pooling and Upsampling layers in U-Net significantly improved T2 and PD map estimation. DeepEMC-T2 mapping aims to estimate one T2 value for each pixel location. As a result, the inclusion of too much unrelated information from distant pixel locations may reduce the accuracy of estimated

T2 and PD values at that specific location in the standard U-Net with Pooling and Unsampling layers. Third, the use of spatiotemporal convolution kernels helps extract additional temporal correlations along the echo dimension compared to spatial convolution kernels only, which contributes to the improvement in the accuracy of the estimated T2 and PD maps.

**Impact of Echo Spacing**

Increasing the echo spacing while simultaneously reducing the number of echoes (e.g., from 5 to 3 echoes) has minimal impact on the accuracy of T2 and PD estimation for DeepEMC-T2 mapping. Meanwhile, maintaining the same number of echoes while expanding the echo spacing proves beneficial in improving the accuracy of T2 and PD estimation for DeepEMC-T2 mapping. This is likely because of the increased T2 decay range that contributes to a more accurate estimation of the T2 and PD maps for DeepEMC-T2 mapping. This finding suggests that it is possible to acquire fewer echoes with a larger echo spacing, and this can be leveraged to reduce SAR without affecting the accuracy of the parameter estimation.

**Importance of Spatial Information**

Compared to the FCN, DeepEMC-T2 mapping with our modified U-Net demonstrates a significant improvement in the estimation of T2 and PD maps. The improvement can be attributed to three factors. First, a simple FCN model is unable to effectively approximate the complex and large dictionary, leading to increased error in the estimation of T2 and PD values. Secondly, training the model at the image level allows for the capture of spatial correlations within MESE images, which proves more beneficial when the number of echoes is reduced. Third, training a model on temporal curves only is susceptible to the influence of noise present in the reconstructed MESE images. This makes it less robust when compared to using the entire image as input.

**Extension of the Study**

This study has several limitations that require discussion. First, the current implementation of DeepEMC-T2 mapping is based on DICOM images that are reconstructed directly on the scanner. As a result, it is a post-processing step after image reconstruction. It is expected that our DeepEMC-T2 mapping framework can also be implemented with accelerated MESE images, combining image reconstruction and parameter estimation as a single step. This can potentially reduce total scan times and/or improve spatial resolution.

Second, DeepEMC-T2 mapping can be implemented with datasets acquired using a radial MESE sequence. In particularly, radial MESE imaging with a golden-angle radial rotation scheme (32,33) could be particularly interesting for DeepEMC-T2 mapping. This may allow for a higher acceleration of data acquisition, leveraging the incoherent sampling behavior of radial sampling. It can also enable improved motion robustness for applications in the spinal cord or abdominopelvic organs.

Third, the DeepEMC-T2 mapping framework was trained using datasets acquired with the same imaging parameters (TR, TE, etc.) and the same acquisition orientation (e.g., axial plane). Our testing datasets have imaging parameters and acquisition orientation which are matched to the training datasets to ensure robust parameter estimation. However, this may limit the use of this network for new datasets acquired with different imaging parameters or orientations. One possible solution to address this challenge is to include more datasets acquired with different imaging parameters for training to improve the generalizability of the network. This will be explored in future work.

Fourth, while this study only focused on a single compartmental model, recent studies have shown that multicomponent analysis can improve myelin water imaging and the investigation of microstructural compartmentation in general (18). However, the challenges associated with multicomponent analysis include limitations in computational power, particularly as the size of the dictionary increases. These can also be addressed with an extension of DeepEMC-T2 mapping to a multi-compartmental model.

## Conclusion

This work introduces a novel deep learning framework to implement EMC-T2 mapping. It addresses challenges associated with standard EMC-T2: the need for a time-consuming dictionary matching step and an adequate number of echoes in MESE imaging. DeepEMC-T2 mapping enables simple, efficient, and accurate T2 quantification directly from acquired MESE images with a reduced number of echoes. This allows one to reduce the number of 180$^o$ refocusing pulses, leading to an important increase in coverage to achieve whole brain imaging with more clinically feasible scan times. The new DeepEMC-T2 framework could facilitate more widespread clinical translation of the EMC-T2 mapping technique for various applications and also more efficient and robust multicomponent analysis (18).

## Acknowledgment

This work was supported by the NIH (R01EB030549, R21EB032917, and P41EB017183) and was performed under the rubric of the Center for Advanced Imaging Innovation and Research (CAI$^2$R), an NIBIB National Center for Biomedical Imaging and Bioengineering. The authors thank Michelle Ng for help with preparing the training datasets.

## References

1. Poon, Colin S., and R. Mark Henkelman. "Practical T2 quantitation for clinical applications." Journal of Magnetic Resonance Imaging 2.5 (1992): 541-553.
2. Bauer, Sonja, et al. "Quantitative T2′-mapping in acute ischemic stroke." Stroke 45.11 (2014): 3280-3286.
3. O'Brien, Aaron T., et al. "T2 mapping in myocardial disease: a comprehensive review." Journal of Cardiovascular Magnetic Resonance 24.1 (2022): 33.
4. Feng, Li, et al. "Accelerated cardiac T2 mapping using breath‐hold multiecho fast spin‐echo pulse sequence with k‐t FOCUSS." Magnetic resonance in medicine 65.6 (2011): 1661-1669.
5. Surowiec, Rachel K., Erin P. Lucas, and Charles P. Ho. "Quantitative MRI in the evaluation of articular cartilage health: reproducibility and variability with a focus on T2 mapping." Knee Surgery, Sports Traumatology, Arthroscopy 22 (2014): 1385-1395.
6. Soellner, S. T., et al. "Intraoperative validation of quantitative T2 mapping in patients with articular cartilage lesions of the knee." Osteoarthritis and cartilage

25.11 (2017): 1841-1849.

7. Mai, Julia, et al. "T2 mapping in prostate cancer." Investigative radiology 54.3 (2019): 146-152.
8. Chatterjee, Aritrick, et al. "Performance of T2 maps in the detection of prostate cancer." Academic radiology 26.1 (2019): 15-21.
9. Hepp, Tobias, et al. "T2 mapping for the characterization of prostate lesions." World Journal of Urology 40.6 (2022): 1455-1461.
10. Cieszanowski, Andrzej, et al. "Characterization of focal liver lesions using quantitative techniques: comparison of apparent diffusion coefficient values and T2 relaxation times." European radiology 22 (2012): 2514-2524.
11. Patten, Carolynn, Ronald A. Meyer, and James L. Fleckenstein. "T2 mapping of muscle." Seminars in musculoskeletal radiology. Vol. 7. No. 04. Copyright© 2002 by Thieme Medical Publishers, Inc., 333 Seventh Avenue, New York, NY 10001, USA.
12. Ben-Eliezer, Noam, Daniel K. Sodickson, and Kai Tobias Block. "Rapid and accurate T2 mapping from multi–spin-echo data using Bloch-simulation-based reconstruction." Magnetic resonance in medicine 73.2 (2015): 809-817.
13. Nassar, Jannette, et al. "Estimation of subvoxel fat infiltration in neurodegenerative muscle disorders using quantitative multi-T2 analysis." NMR in Biomedicine 36.9 (2023): e4947.
14. Holodov, Maria, et al. "Probing muscle recovery following downhill running using precise mapping of MRI T 2 relaxation times." Magnetic Resonance in Medicine 90.5 (2023): 1990-2000.
15. Chechik, Yigal, et al. "Post-run T2 mapping changes in knees of adolescent basketball players." Cartilage 13.1_suppl (2021): 707S-717S.
16. Solomon, Chen, et al. "Psychophysical evaluation of visual vs. computer-aided detection of brain lesions on magnetic resonance images." Journal of Magnetic Resonance Imaging 58.2 (2023): 642-649.
17. Shepherd, Timothy M., et al. "New rapid, accurate T2 quantification detects pathology in normal-appearing brain regions of relapsing-remitting MS patients." NeuroImage: Clinical 14 (2017): 363-370.

18. Omer, Noam, et al. "Data‐driven algorithm for myelin water imaging: probing subvoxel compartmentation based on identification of spatially global tissue features." Magnetic resonance in medicine 87.5 (2022): 2521-2535.
19. Shpringer, Guy, David Bendahan, and Noam Ben-Eliezer. "Accelerated reconstruction of dictionary-based T2 relaxation maps based on dictionary compression and gradient descent search algorithms." Magnetic Resonance Imaging 87 (2022): 56-66.
20. Radunsky, Dvir, et al. "Quantitative platform for accurate and reproducible assessment of transverse (T2) relaxation time." NMR in Biomedicine 34.8 (2021): e4537.
21. Ronneberger, Olaf, Philipp Fischer, and Thomas Brox. "U-net: Convolutional networks for biomedical image segmentation." Medical Image Computing and Computer-Assisted Intervention–MICCAI 2015: 18th International Conference, Munich, Germany, October 5-9, 2015, Proceedings, Part III 18. Springer International Publishing, 2015.
22. Kingma DP, Ba JL. Adam: A Method for Stochastic Optimization. 3rd Int. Conf. Learn. Represent. ICLR 2015 - Conf. Track Proc. 2014 doi: 10.48550/arxiv.1412.6980.
23. Cohen, Ouri, Bo Zhu, and Matthew S. Rosen. "MR fingerprinting deep reconstruction network (DRONE)." Magnetic resonance in medicine 80.3 (2018): 885-894.
24. Feng, Li, Dan Ma, and Fang Liu. "Rapid MR relaxometry using deep learning: An overview of current techniques and emerging trends." NMR in Biomedicine 35.4 (2022): e4416.
25. Liu, Fang, Li Feng, and Richard Kijowski. "MANTIS: model‐augmented neural network with incoherent k‐space sampling for efficient MR parameter mapping." Magnetic resonance in medicine 82.1 (2019): 174-188.
26. Liu, Fang, et al. "Magnetic resonance parameter mapping using model‐guided self‐supervised deep learning." Magnetic resonance in medicine 85.6 (2021): 3211-3226.

27. Liu, Fang, et al. "High-performance rapid MR parameter mapping using model-based deep adversarial learning." Magnetic resonance imaging 74 (2020): 152-160.
28. Pei, Haoyang, et al. "Rapid 3D T1 mapping using deep learning-assisted Look-Locker inversion recovery MRI." Magnetic Resonance in Medicine 90.2 (2023): 569-582.
29. Jun, Yohan, et al. "Deep model-based magnetic resonance parameter mapping network (DOPAMINE) for fast T1 mapping using variable flip angle method." Medical Image Analysis 70 (2021): 102017.
30. Qiu, Shihan, et al. "Multiparametric mapping in the brain from conventional contrast-weighted images using deep learning." Magnetic resonance in medicine 87.1 (2022): 488-495.
31. Sun, Haoran, et al. "Retrospective T2 quantification from conventional weighted MRI of the prostate based on deep learning." Frontiers in Radiology 3 (2023): 1223377.
32. Feng, Li. "Golden-angle radial MRI: basics, advances, and applications." Journal of Magnetic Resonance Imaging 56.1 (2022): 45-62.
33. Feng, Li, et al. "Golden-angle radial sparse parallel MRI: combination of compressed sensing, parallel imaging, and golden-angle radial sampling for fast and flexible dynamic volumetric MRI." Magnetic resonance in medicine 72.3 (2014): 707-717.

## Figure Legend

### Figure 1

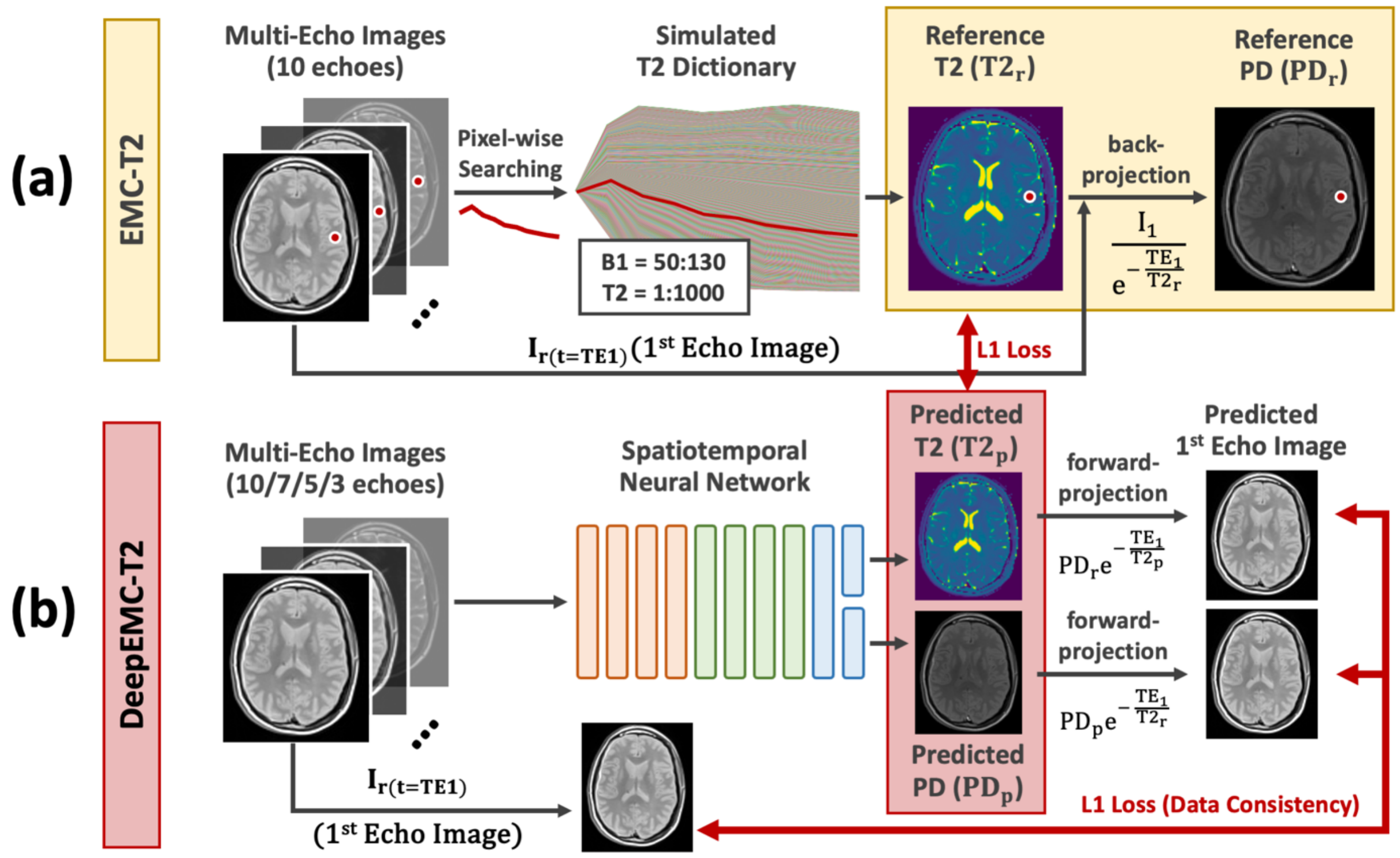

(a) Standard EMC-T2 framework: A T2 map is generated through pixel-wise T2 dictionary matching and a PD map is generated by back-projecting the first echo image with the estimated T2 map. (b) New DeepEMC-T2 framework: A spatiotemporal U-net is developed and trained on MESE images with variable echoes using a supervised scheme. The reference T2 and PD maps for training are calculated from images with 10 echoes using the standard EMC-T2 mapping approach. The original MESE images with 10 echoes were retrospectively cut to 7, 5, and 3 echoes for evaluation, which can potentially enable increased volumetric coverage and/or reduced SAR.

**Figure 2**

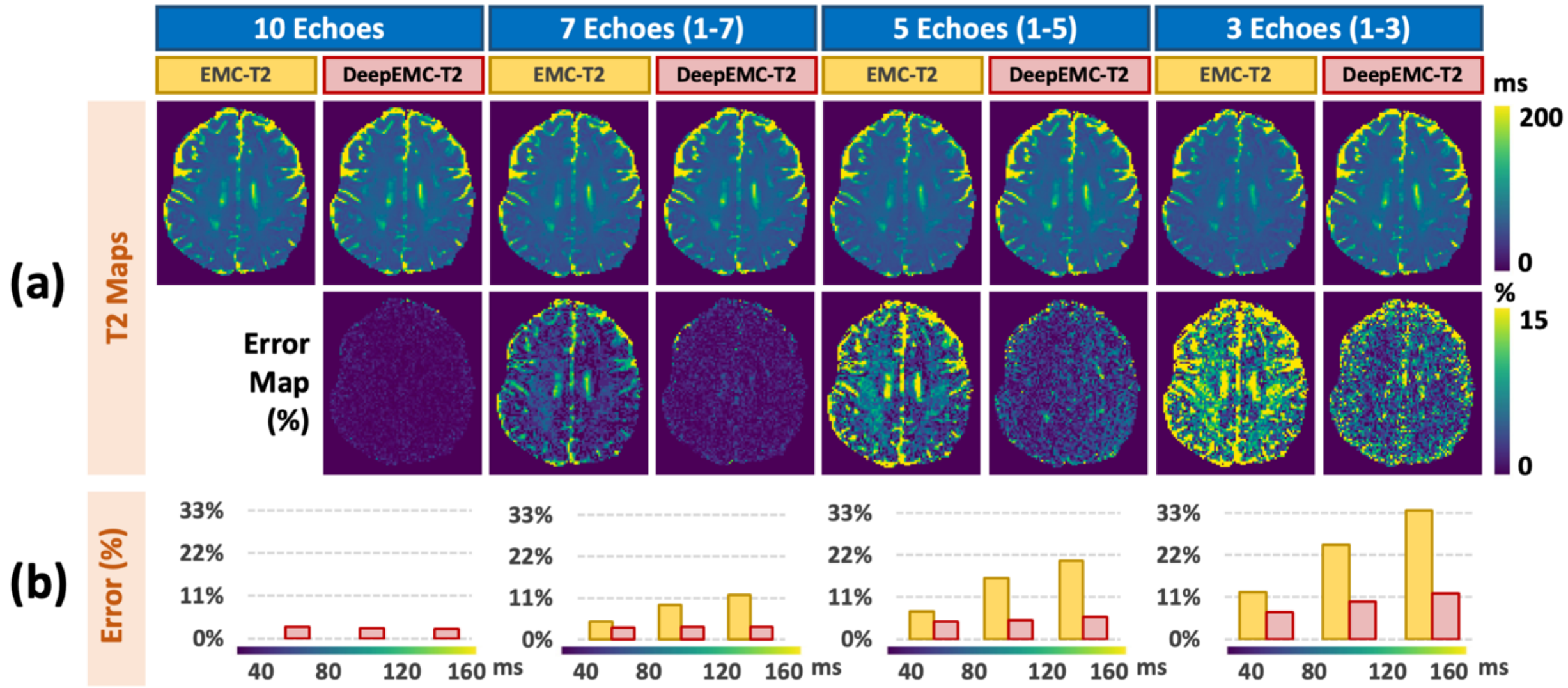

(a) A representative case comparing T2 maps estimated using EMC-T2 and DeepEMC-T2 from MESE images with varying numbers of echoes. The T2 maps estimated from standard EMC-T2 using all 10 echoes were treated as the reference standard for calculating the error maps. (b) Quantitative comparison of T2 estimation in all the testing cases with varying numbers of echoes based on averaged pixel-wise errors in different T2 ranges. The error map and quantitative comparison indicate that DeepEMC-T2 mapping enables more accurate T2 map estimation than EMC-T2 mapping, particularly as the number of echoes is reduced.

**Figure 3**

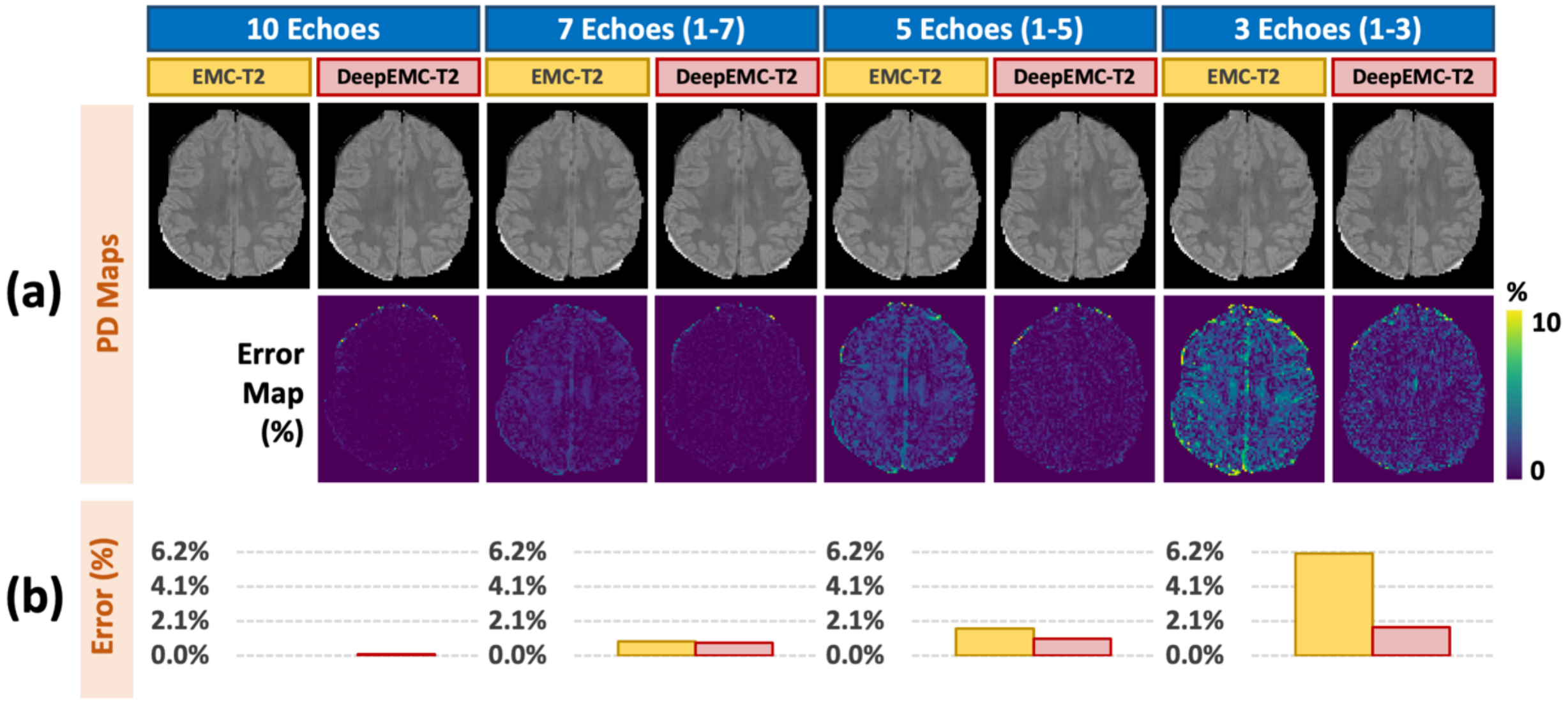

(a) A representative case comparing PD map estimated using EMC-T2 and DeepEMC-T2 from MESE images with varying numbers of echoes. The PD maps estimated from standard EMC-T2 using all 10 echoes were treated as the reference for calculating the error maps. (b) Quantitative comparison of PD estimation in all the testing cases with varying numbers of echoes with different numbers of echoes based on averaged pixel-wise errors. The error map and quantitative comparison indicate that DeepEMC-T2 mapping enables more accurate PD map estimation than EMC-T2 mapping, particularly as the number of echoes is reduced.

**Figure 4**

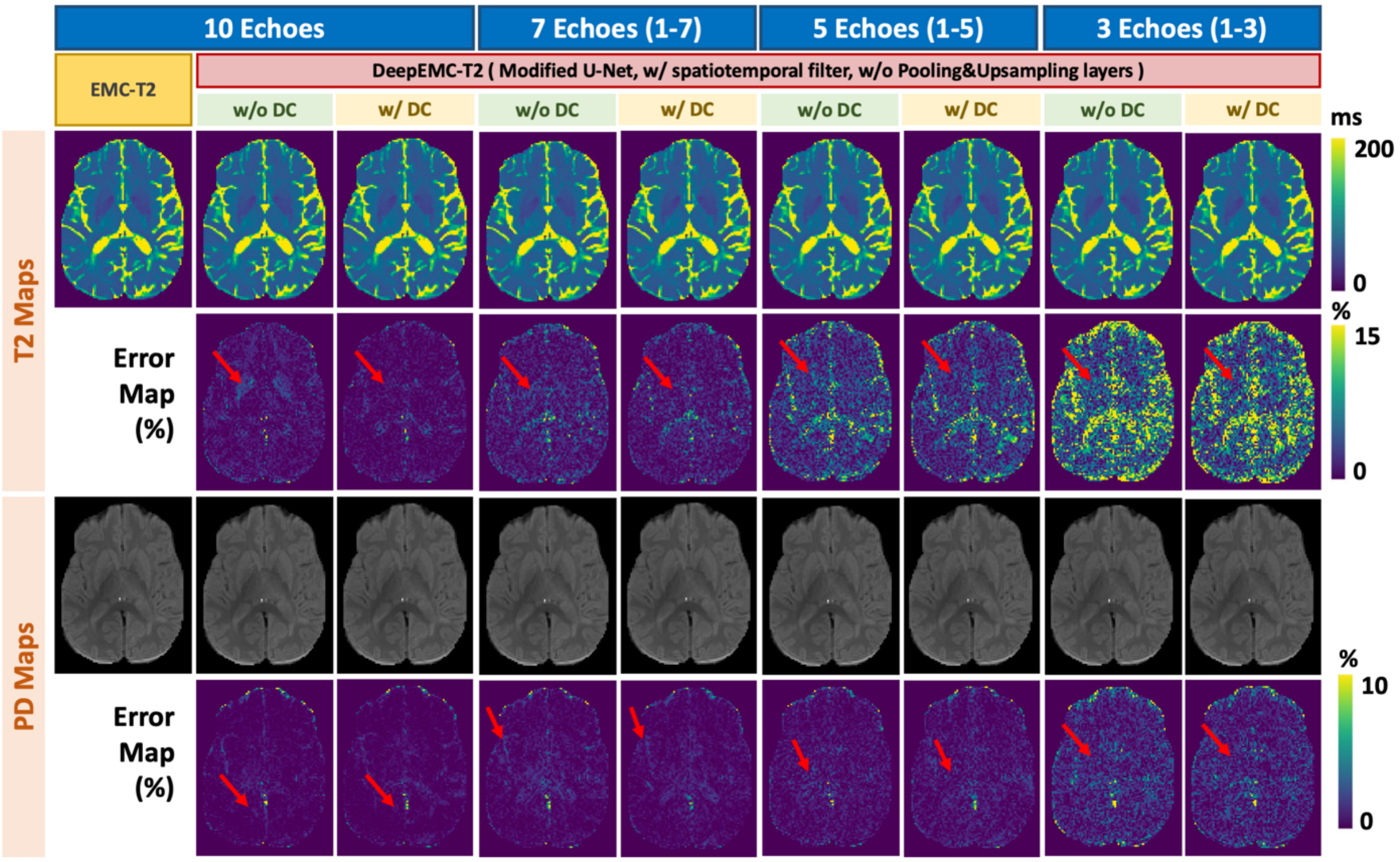

A representative case comparing T2, and PD maps estimated from MESE images using DeepEMC-T2 with and without DC loss. The error maps indicate that incorporating a DC loss into the training of the DeepEMC-T2 can improve the estimation of both T2 and PD maps, as highlighted by the arrows in the error maps. However, the visual impact is relatively small.

**Figure 5**

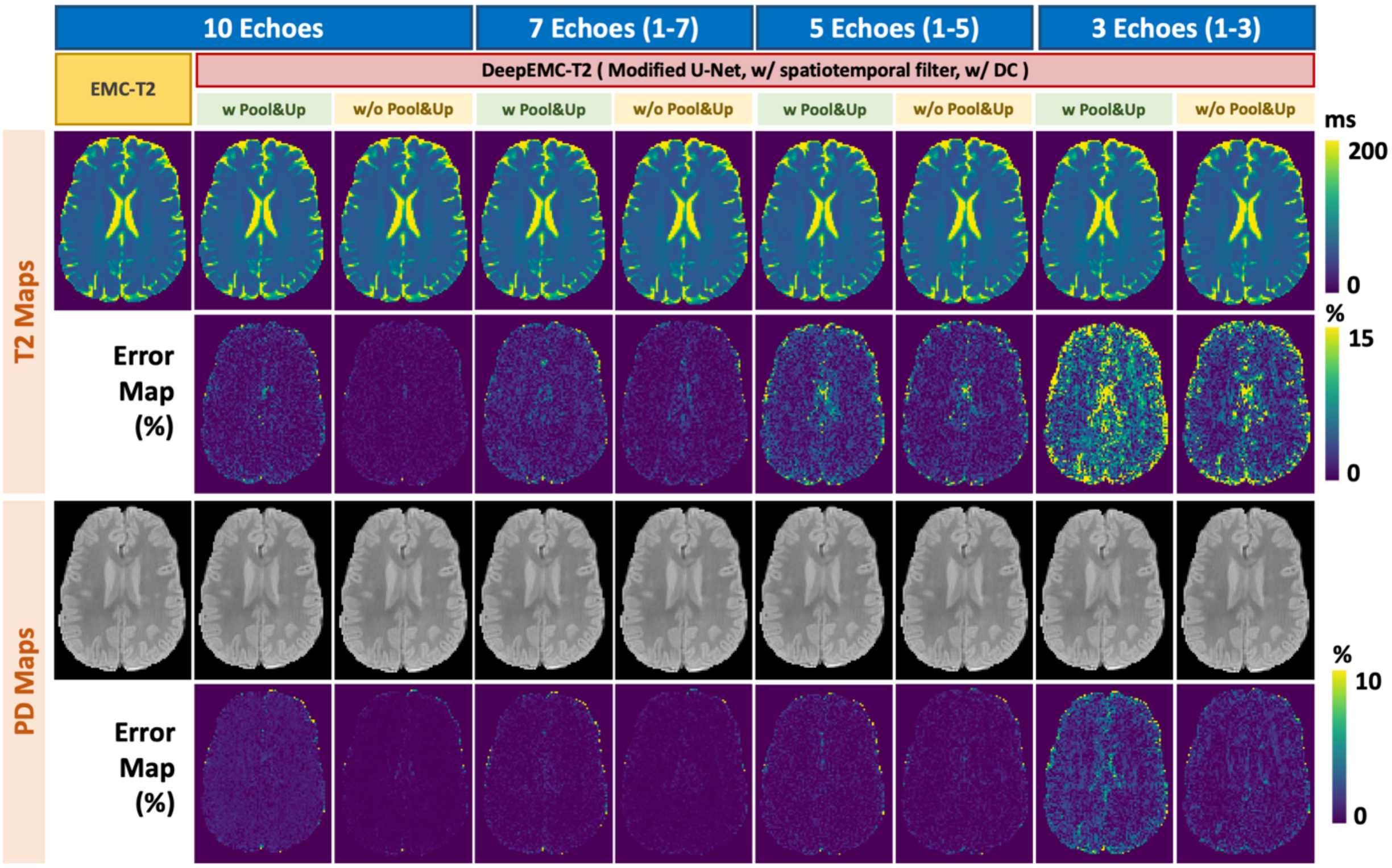

A representative case comparing T2, and PD maps estimated using DeepEMC-T2 with and without the Pooling and Upsampling layers. The error maps indicate a significant reduction in error for both T2 and PD estimation by removing the Pooling and Upsampling layers.

**Figure 6**

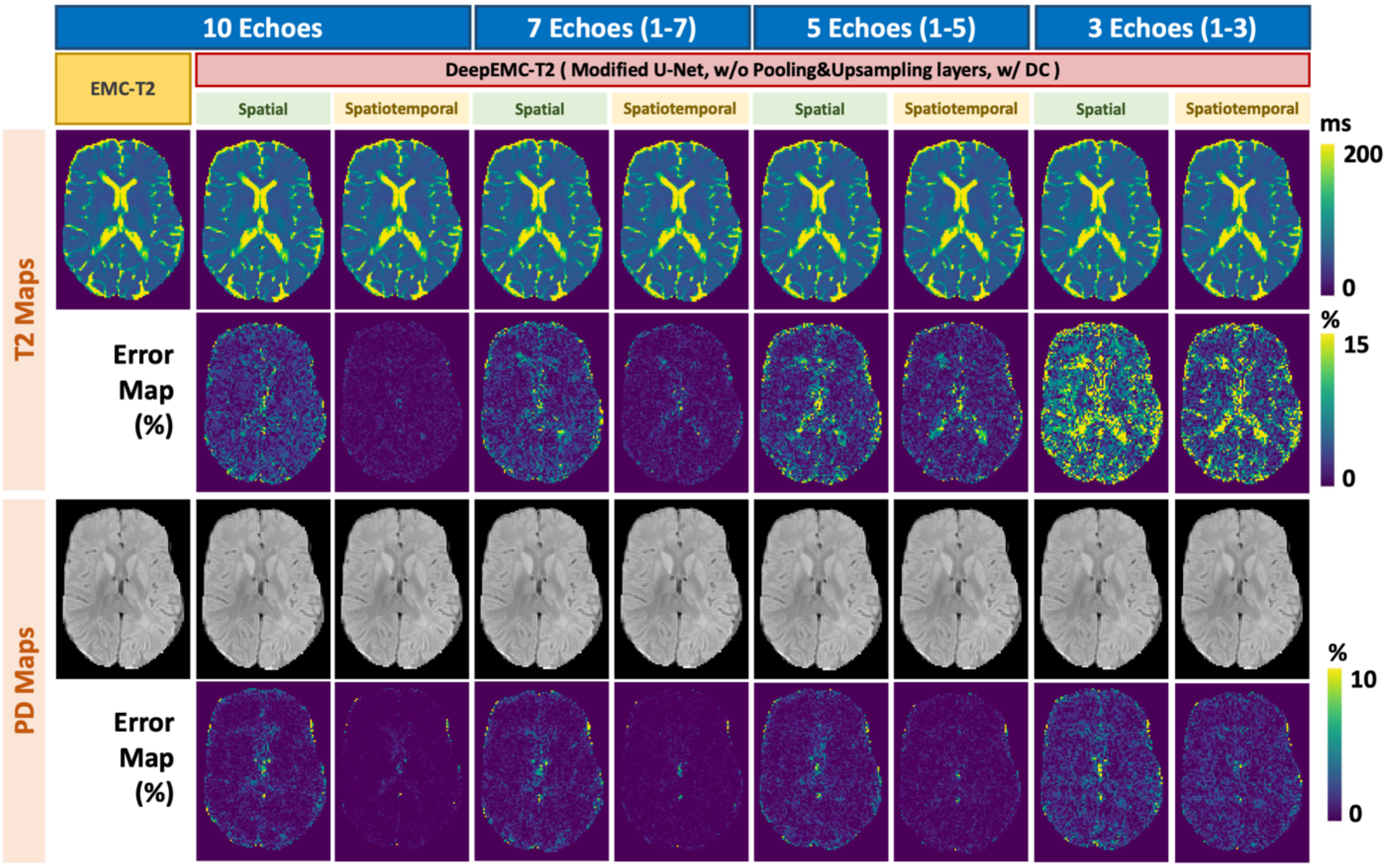

A case comparing T2, and PD maps estimated using DeepEMC-T2 with spatial and spatiotemporal convolution kernels in one case. Error maps suggest that the use of spatiotemporal convolution filters in DeepEMC-T2 results in a significant improvement of T2 and PD estimation.

**Figure 7**

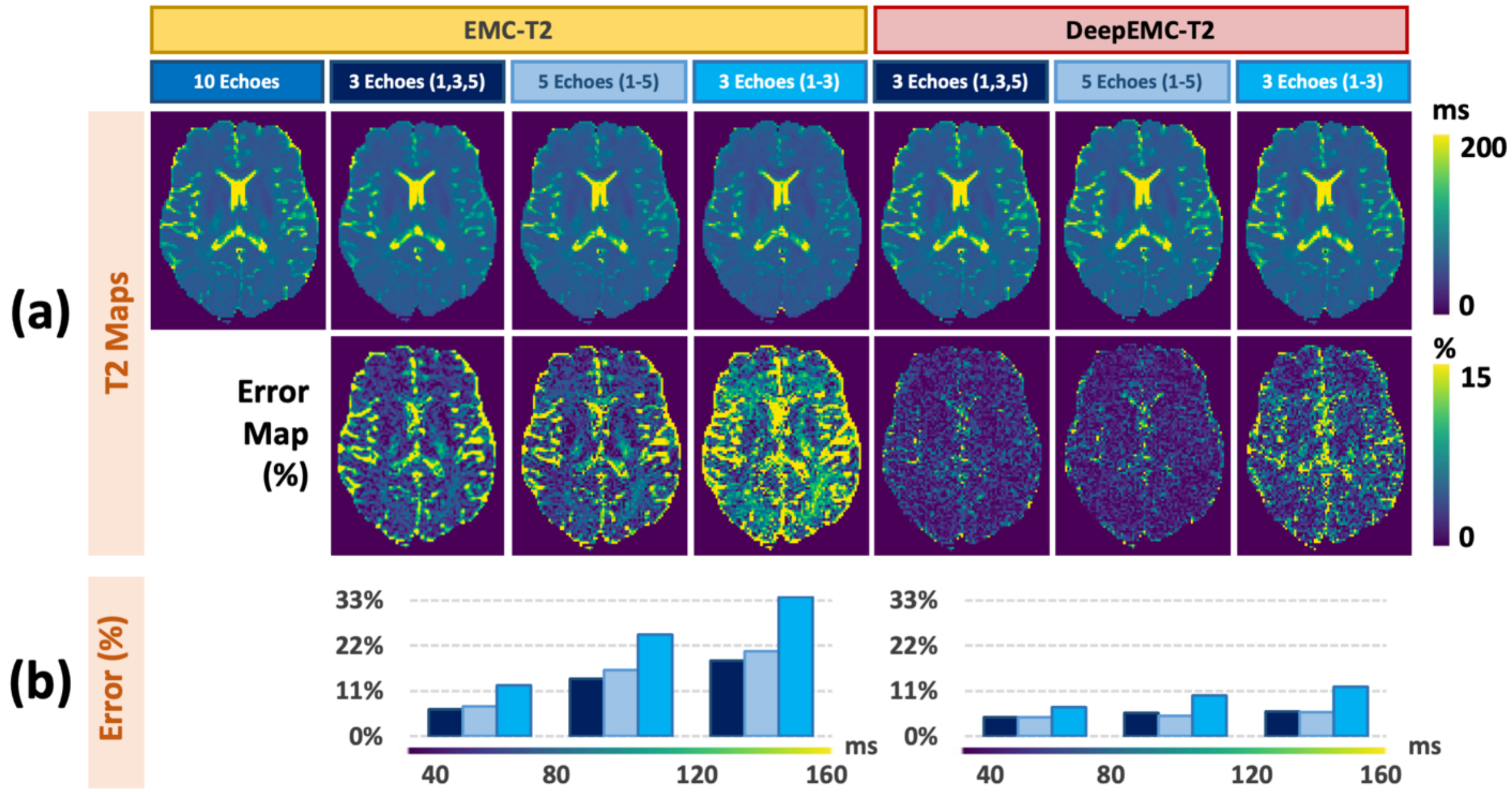

(a) A representative example comparing T2 maps estimated from MESE images for echo 1,3,5 with a larger echo spacing, and for 5 echoes (echo 1-5) and 3 echoes (echo 1-3) with the default echo spacing using both EMC-T2 and DeepEMC-T2. (b) Quantitative comparison of T2 maps in testing datasets based on averaged pixel-wise errors in different T2 ranges. The results suggest that increasing the echo spacing while reducing the number of echoes has little impact on the accuracy of T2 map estimation in DeepEMC-T2 mapping.

**Figure 8**

(a)

(b)

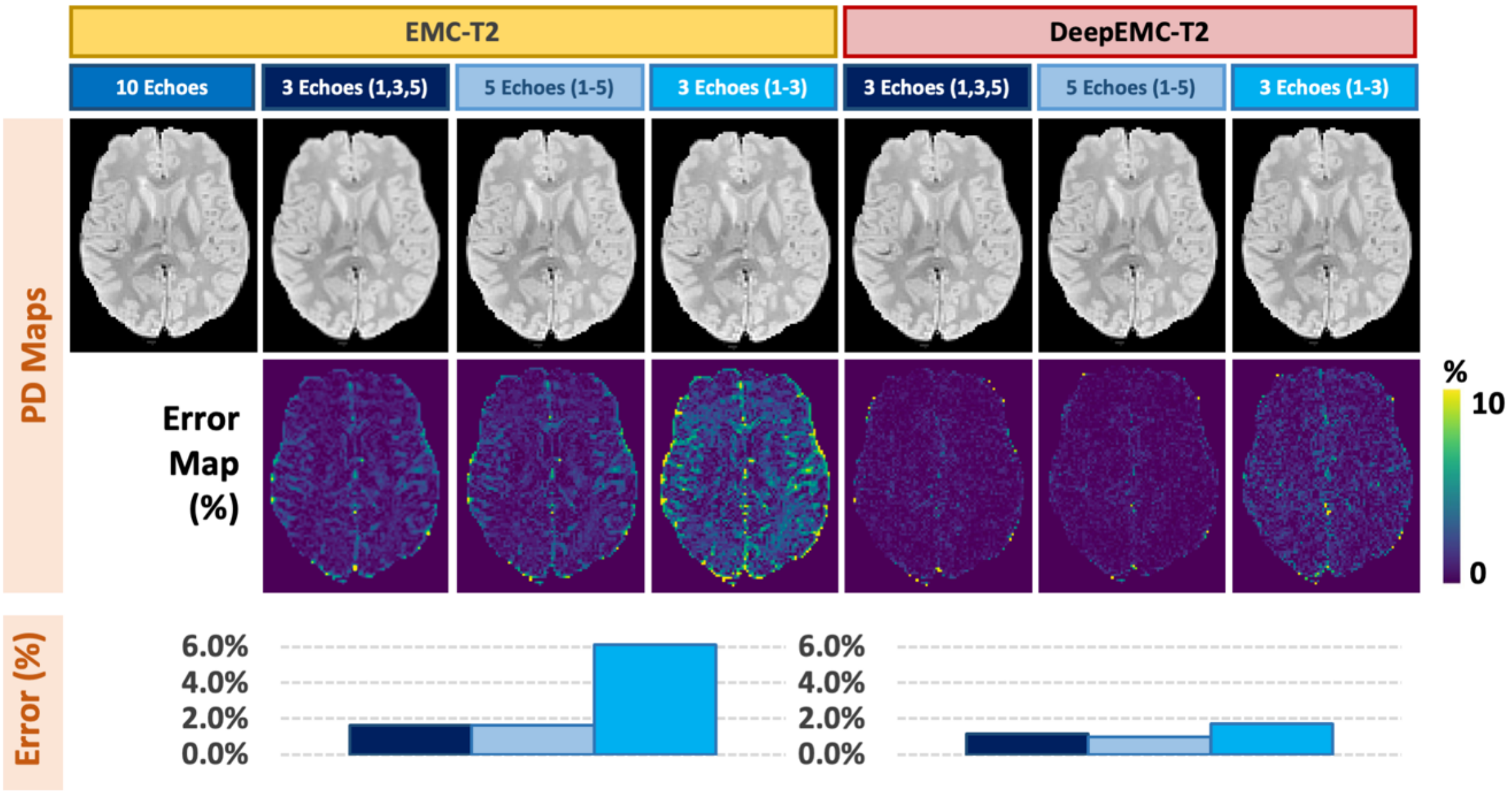

(a) A representative example comparing PD maps estimated from MESE images for echo 1,3,5 with a larger echo spacing, and for 5 echoes (echo 1-5) and 3 echoes (echo 1-3) with the default echo spacing using both EMC-T2 and DeepEMC-T2. (b) Quantitative comparison of PD maps estimated in 15 testing datasets based on averaged pixel-wise errors. The results demonstrated that increasing the echo space while reducing the number of echoes has little impact on the accuracy of PD map estimation in DeepEMC-T2 mapping.

**Figure 9**

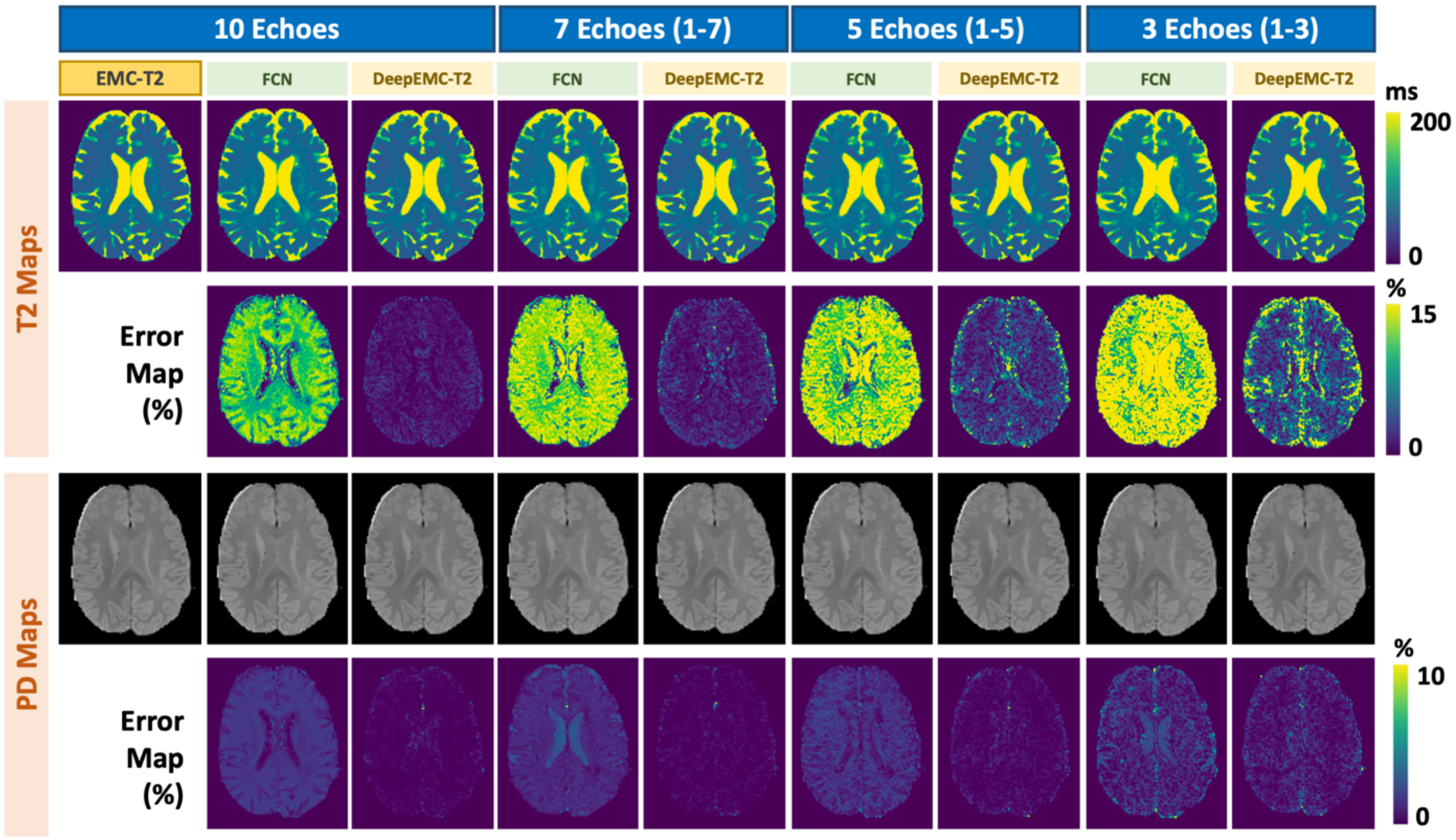

A representative example comparing T2 and PD maps estimated using modified U-Net and FCN in the DeepEMC-T2 for MESE images with varying numbers of echoes. The results demonstrated that DeepEMC-T2 exhibits better accuracy compared to FCN across different numbers of echoes. This highlights the significance of spatial correlations in estimating parameter maps for DeepEMC-T2.

**Table 1**

| T2 ranges (ms) / PD | 40-80 | 80-120 | 120-160 | PD | 40-80 | 80-120 | 120-160 | PD | 40-80 | 80-120 | 120-160 | PD | 40-80 | 80-120 | 120-160 | PD |
|---|---|---|---|---|---|---|---|---|---|---|---|---|---|---|---|---|
| Network Architecture | Error (%) | | | | | | | | | | | | | | | |
| | 10 Echoes | | | | 7 Echoes (1-7) | | | | 5 Echoes (1-5) | | | | 3 Echoes (1-3) | | | |
| ✓ DC Loss<br>✓ Spatiotemporal<br>✗ Pool&Up layers<br>**(Proposed)** | **2.93** | **2.59** | **2.39** | 0.73 | **3.12** | **3.40** | **3.38** | **0.75** | 4.68 | **4.91** | 5.85 | **1.00** | **7.00** | **9.86** | 12.01 | **1.69** |
| ✗ DC Loss<br>✓ Spatiotemporal<br>✗ Pool&Up layers | 3.20 | 3.01 | 2.78 | **0.72** | 3.73* | 3.70 | 3.55 | 0.86* | **4.50** | 5.16 | **5.78** | 1.04 | 7.69* | 10.08 | **11.91** | 1.81* |
| ✓ DC Loss<br>✓ Spatiotemporal<br>✓ Pool&Up layers | 3.92* | 3.52* | 3.70* | 1.12* | 4.42* | 4.38* | 4.52* | 1.22* | 5.29* | 6.14* | 6.76* | 1.50* | 10.11* | 13.46* | 16.34* | 3.22* |
| ✓ DC Loss<br>✓ Spatial<br>✗ Pool&Up layers | 5.17* | 5.31* | 5.79* | 1.68* | 4.65* | 4.98* | 5.75* | 1.53* | 6.02* | 7.05* | 8.00* | 1.72* | 8.01* | 11.38* | 13.85* | 2.34* |
| Fully Connected Network (FCN) | 7.72* | 10.07* | 9.34* | 0.76 | 8.27* | 10.86* | 10.43* | 1.05* | 8.98* | 12.48* | 11.17* | 1.06 | 13.09* | 20.34* | 22.51* | 2.08* |

Quantitative comparison of T2 and PD maps estimated from MESE images using various new features and fully connected network (FCN) of DeepEMC-T2 across all testing datasets based on averaged pixel-wise errors. The results confirm that incorporating DC loss, removing Pooling and Upsampling layers, and using spatiotemporal convolution kernels all lead to the improvement in the estimation of T2 and PD maps. The proposed DeepEMC-T2 model has better performance in estimating T2 and PD maps compared to the FCN model.

**Figure S1**

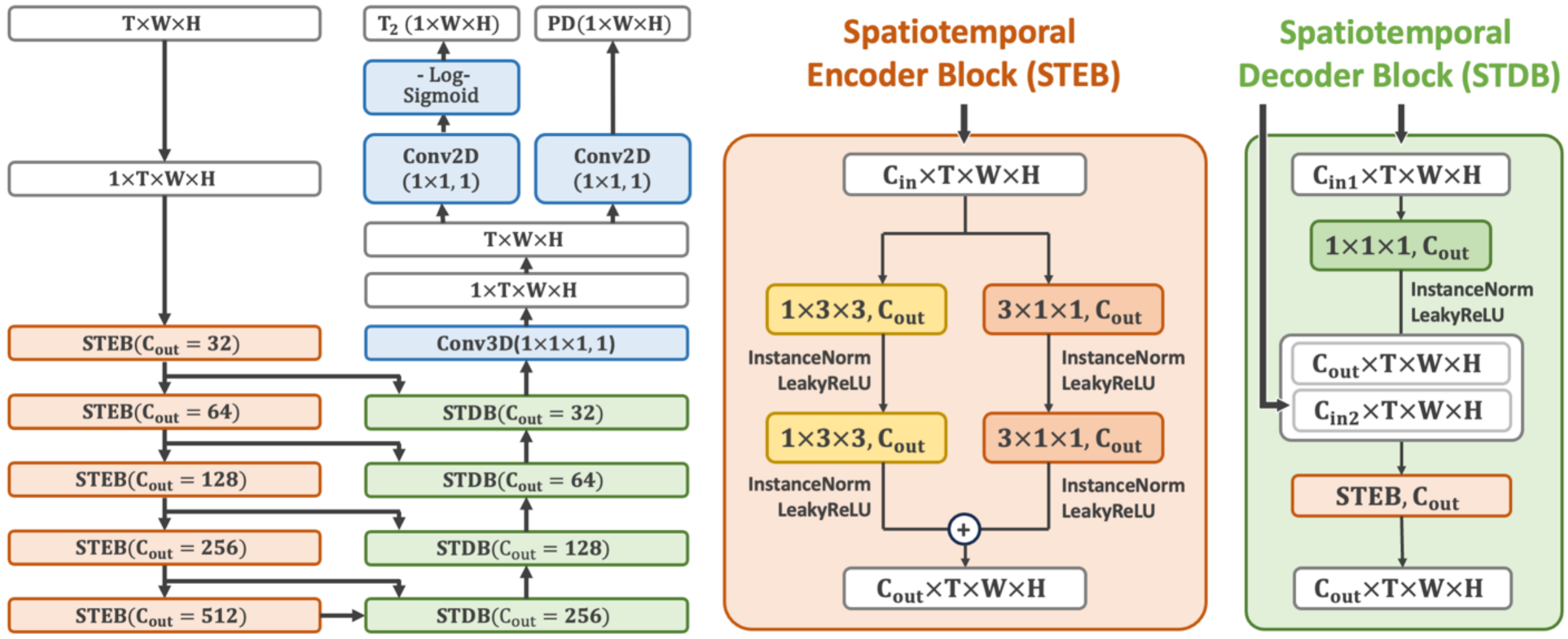

The detailed architecture of the DeepEMC-T2 modified from a standard U-Net. First, the spatial convolution filters in the standard U-Net are replaced with spatiotemporal convolution filters to efficiently extract the temporal correlations between different echo images. Second, Pooling and Upsampling layers that are implemented in the standard U-net are removed to perform pixel-wise parameter generation. The linear transforms were employed at the final layer to map the features at each pixel to T2 and PD maps following all spatiotemporal encoder and decoder blocks, which was implemented using two 2D convolutional layers (Conv2D) with a kernel size of $1 \times 1$.

**Figure S2**

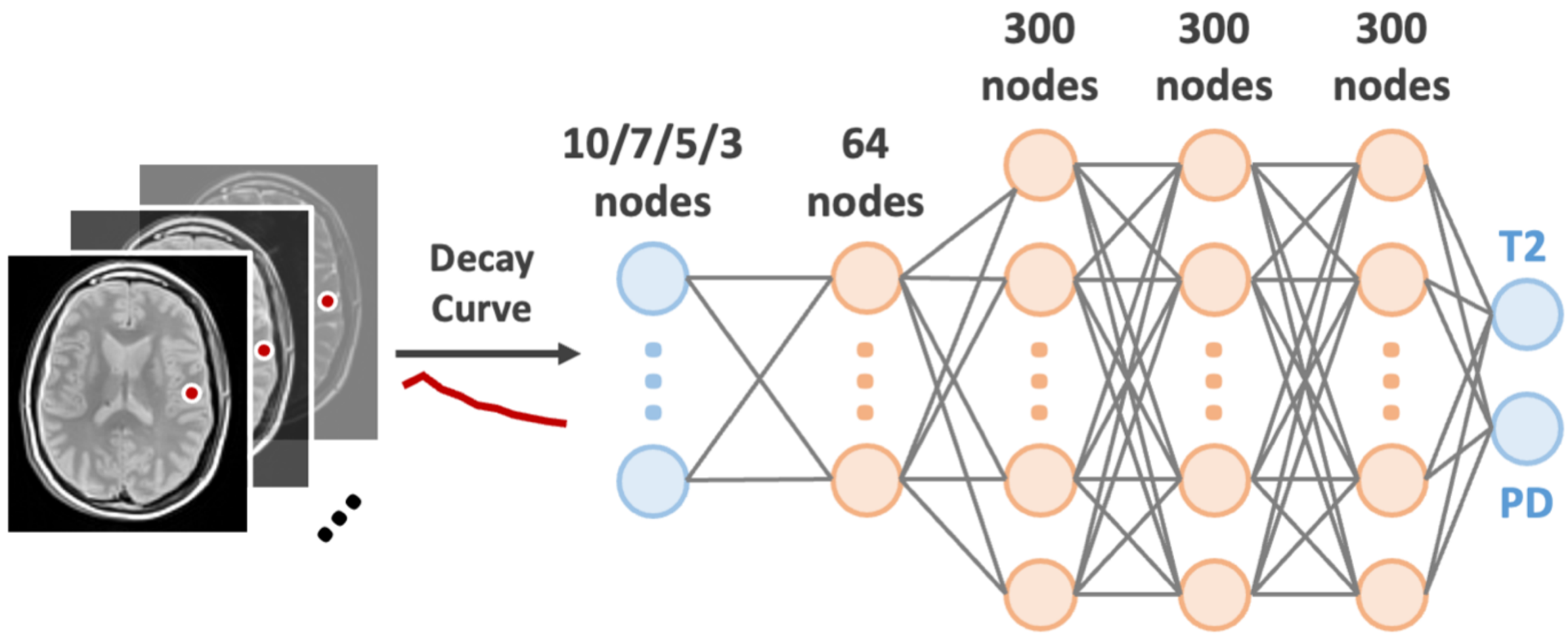

The detailed architecture of fully connected network. The FCN was only trained on the decay curve at each pixel location of MESE images to estimate the T2 and PD values at corresponding pixel location. Importantly, the FCN did not take the entire MESE images as input. Consequently, the estimation of T2 and PD maps did not rely on any spatial information or features. The hidden layer of the FCN utilizes ReLU as the activation function. Additionally, the activation function of the output node for predicting T2 values was modified to a negative Log-Sigmoid.